\documentclass[12pt]{article}
\usepackage{graphicx}
\usepackage{amssymb,amsmath,amsfonts,palatino,amsthm}
\usepackage{amssymb}
\usepackage{epstopdf}
\theoremstyle{definition}

\newcommand{\beq}{\begin{equation}}
\newcommand{\eeq}{\end{equation}}

\DeclareMathOperator{\tr}{tr}

\newcommand{\A}{\mathcal{A}}
\newcommand{\Hi}{\mathcal{H}}

\newcommand{\f}{\begin{equation}}
\newcommand{\ff}{\end{equation}}

\newcommand{\schwarz}{\text{Schwarzschild}} 
\newcommand{\hschw}{\Hi_A^{\scriptsize \schwarz}} 
\newcommand{\hall}{\Hi_A^{\scriptsize\text{All}}} 
\newcommand{\hinter}{H^{\scriptsize\text{int}}}
\newcommand{\ainter}{\A^{\scriptsize\text{int}}}
\newcommand{\aginter}{\A^{\scriptsize\text{gint}}}
\newcommand{\cainter}{{\A^{\scriptsize\text{int}}}'}
\newcommand{\hsurf}{\Hi^{\mathcal{S}}}
\newcommand{\hjs}{{\cal V}_{\{ j \}}}

\begin{document} \title{Symmetry and entropy of black hole horizons}
\author{Olaf Dreyer\thanks{Email address: odreyer@perimeterinstitute.ca},  
 Fotini Markopoulou\thanks{Email address:
fotini@perimeterinstitute.ca}   and Lee Smolin \thanks{Email
address:lsmolin@perimeterinstitute.ca}\\
\\
Perimeter Institute for Theoretical Physics,\\
35 King Street North, Waterloo, Ontario N2J 2W9, Canada, and \\
Department of Physics, University of Waterloo,\\
Waterloo, Ontario N2L 3G1, Canada\\
\\
}
\date{\today}
\maketitle
\vfill
\begin{abstract}
We argue, using methods taken from the theory of noiseless subsystems in
quantum information theory, that the quantum states associated with a Schwarzchild
black hole live in the restricted subspace of the Hilbert space of
horizon boundary states in which all punctures are equal.  Consequently,
one value of the Immirzi parameter matches both the Hawking value for the
entropy and the quasi normal mode spectrum of the 
Schwarzchild black hole.  The method of noiseless subsystems allows us
to understand, in this example and more generally, how symmetries, which
take physical states to physical states, can emerge from a diffeomorphism invariant
formulation of quantum gravity.  
\end{abstract}
\vfill
\newpage

\section{Introduction}

This paper is concerned with two distinct problems in background 
independent formulations of quantum gravity. First, how one can 
recover  the semiclassical Bekenstein-Hawking results\cite{bek1,hawking}
in the context of results about the entropy of horizons and surfaces
in loop quantum gravity\cite{linking}-\cite{olaf-bh}.  
Second, how the description of spacetime in terms of 
classical general relativity emerges from the quantum theory.
Up till now, these problems have been treated in isolation.
Here we propose  a perspective that relates them, and also addresses
a technical issue concerning the black hole entropy. 

This  technical issue concerns the value of a certain free
parameter, $\gamma$,  in the theory called the Immirzi parameter \cite{imirzi}.  This parameter, not present in the classical
theory, arises from an ambiguity in the quantization procedure because there
is a choice in the connection variable that is used in the contruction
of the quantum obserable algebra. As it does not appear in the classical
theory, its value should be fixed by some physical requirement.  
In previous work  \cite{linking}-\cite{isolated}, it was shown that
$\gamma$ is fixed by the requirement that  the quantum
gravity computation reproduce the Hawking value of the entropy.

Since the story is very simple and physical, we review it for non-experts. The Immirzi parameter comes into the formula for the area of a surface. When a surface is punctured by a set of spin network edges with spin labels $\{j_\alpha\}$, then the area of the surface is given by
\f
A [\{ j_\alpha \} ] = 8\pi  \gamma \sum_\alpha \sqrt{j_\alpha (j_\alpha +1)},
\ff
which we take here to be measured in units of the Planck area $l_{Pl}^2$.

In loop quantum gravity, there is an exact description of the quantum
geometry of black hole horizons\cite{isolated}, as well as a more
general class of boundaries\cite{linking}.  The entropy of
the horizon is defined in terms of the Hilbert space of the boundary theory. For
reasons that we spell out shortly, it is taken to be 
the logarithm of the dimension of the Hilbert space of boundary states. 

The Immirzi parameter does not come into the entropy. 
Hence it appears  in the ratio of the entropy $S[A]$ of a black hole
to its area:
\f
{\cal R} = \frac{S[A] }{A [\{ j_\alpha \} ]} =
\frac{c}{\gamma},
\ff
where $c$ is a constant to be determined by calculating the entropy.
At the same time, we believe from Hawking's semiclassical calculation
of black hole entropy that\cite{hawking}
\f
{\cal R}= \frac{1}{4}.
\ff
For the two equations to agree, 
\f
\gamma = 4c.
\ff

Recently, one of us, following a clue uncovered by Hod\cite{hod},
described a semiclassical argument which
fixes $\gamma$ by an argument that appears to be independent of
considerations of entropy or radiation but rather makes use of the 
quasinormal mode spectrum\cite{olaf-bh}.  As it is relevent, we repeat here the basic
argument.
 
The quasinormal mode spectrum of the Schwarzchild black hole turns out
to have an asymptotic form
\f
M \omega= \frac{\ln (3)}{8\pi} + \frac{i}{4} \left(n + \frac{1}{2} \right),
\ff
for arbitrarily large integer $n$.  As $n$ goes to infinity, the decay times of the modes goes
to zero in units of the light crossing time of the horizon.  This means that
the excitations involve  more and more local regions of the horizon.  
At the same time, the real part of the frequency goes to the asymptotic
value
\f
\omega_{qnm} = \frac{\ln (3)}{8\pi M }.
\ff
This frequency is then associated with an arbitrarily short lived, and hence
local, excitation of the horizon.   By the correspondence principle, this
must correspond to an energy
\f
\Delta M_{qnm} = \hbar \omega_{qnm} = \frac{\ln (3)\hbar}{8\pi M }.
\ff

In the classical theory, the region of the horizon excited may be arbitrarily
small, but in the quantum theory there is a smallest region that can be
excited, which is a minimal puncture with minimal area
\f
\Delta A= 8\pi \gamma \sqrt{j_{min} (j_{min}+1) }.
\ff
However, for a Schwarzchild black hole,
\f
A= 16 \pi M^2,
\ff
from which it follows that, for a quasinormal mode,
\f
\Delta A_{qnm}= 32 \pi M \Delta M =  4 \ln (3) l_{Pl}^2.
\ff
The result is a prediction for the Immirzi parameter, which is
\f
\gamma^{qnm} = \frac{\ln (3)}{2 \pi\sqrt{j_{min}(j_{min}+1 )}}.
\ff

Does this match the value needed to get ${\cal R}=1/4$?  It turns out
to depend on what we take for the Hilbert space of the black hole
horizon.  By arguments given in \cite{isolated}, we know that the horizon
Hilbert space, for fixed area $A$ in Planck units, is related to
the Hilbert space of the invariant states of $U(1)$ 
Chern-Simons theory on an $S^2$ with punctures $\{ j_\alpha \} $,
denoted by ${\cal V}_{\{ j_\alpha\}}$\footnote{There is an equivalent
description in terms of $SU(2)$ Chern-Simons theory\cite{linking}.}.
The level $k$ is related to the mass of the black hole, and is
assumed here to be large.  
Given a set of punctures $\{ j \}$, 
\f
{\cal V}_{\{ j_\alpha\}} = \text{Inv} \left ( \prod_\alpha {\cal H}_{j_\alpha } \right ) 
\ff
where ${\cal H}_{j }$ is the $2j+1$ dimensional spin space for
spin $j$ and Inv means the Hilbert states contains 
only states with total spin zero. 

We know that the Schwarzchild black hole has definite area $A[\{ j \} ]$
and definite energy:
\f
M(\{  j \} ) = \sqrt{\frac{A(\{  j \} )}{16\pi}}.
\ff
As the area is quantized, we see that the energy is quantized as well. 

We then expect that at least some of the Hilbert spaces ${\cal V}_{\{ j_\alpha\}}$
contain states which correspond to Schwarzchild black holes. 
The state is expected to be thermal, because it is entangled with the states of radiation
that has left the black hole. 
We are, howeve,r ignorant of the exact state so, instead, the black hole entropy is
usually taken to be 
\f
S^{\schwarz} = \text{ln } \text{dim } \hschw,
\ff
where $\hschw$ is a Hilbert space on which the density matrix 
$\rho_A^{\schwarz}$ describing the ensemble of Schwarzschild black holes 
may be expected to be non-degenerate.  

One proposal is to take
\f
\hschw = \hall = \sum_{A(\{ j\} ) = A }  \hjs
\ff
where we use the superscript $\text{All}$ to remind us that in this case we
sum over all sets of punctures that give an area between $A$ and $A+\Delta A$,
for some small $\Delta A$.  

However, this proposal can be criticized on the basis that the only information
about the Schwarzchild black hole that is used is its area.
We may expect that the actual state of the Schwarchild black hole
is determined by additional physical considerations and hence 
is non-degenerate only in a 
subspace 
\f
\hschw \subsetneq\hall
\ff
that is picked out by additional physical input. 

In this article, we will argue that the physical Hilbert
space associated to the horizon of a Schwarzschild black hole is 
dominated by the $\hjs$ for which all punctures have equal minimal spin $j_{\text{min}}$. 
The difference is important, because in the first case we have
\f
{\cal R}^{\text{All}}=\frac{\text{ln } \text{dim } \hall}{A (\{ j \})},
\ff
from which recent calculations\cite{nomatch}  have deduced the value\footnote{But see
\cite{takashi} for a criticism of that calculation.}
\f
\gamma^{\text{All}}=  0.23753295796592\ldots .
\ff
On the other hand, in the second case
\f
{\cal R}^{\text{min}}=\frac{\text{ln }\text{dim }{\cal V}_{\{ j_{min},..., j_{min}\}}  }
{A (\{ j \} )}.
\ff
This is easy to compute \cite{linking,olaf-bh} and leads to a value
\f
\gamma^{\text{min}}=  \frac{\text{ln }(3) }{2\pi \sqrt{j_{\text{min}}(j_{\text{min}}+1 )}}.
\ff
We note that
\f
\gamma^{\text{qnm}}= \gamma^{\text{min}} \neq \gamma^{\text{All}}.
\ff

Hence, we have the following situation:

If indeed the right choice for the Hilbert space of a Schwarzchild
black hole consists only of states with equal and minimal punctures,
then there is a remarkable agreement between the two computations.
This supports the case that there is something physically correct
about the description of black holes in loop quantum gravity.

On the other hand, if $\hall$ is the right Hilbert space
for the horizon of a Schwarz- child black hole, loop quantum gravity is
in deep trouble, because there is no choice of $\gamma$ that will
agree with both the Hawking entropy and the quasi-normal mode
spectrum.

It is clear that to resolve this problem we need additional physical input.
The only property of the Schwarzchild black hole used in previous work
to constrain the corresponding quantum state is its area.  But a more
complete treatement should take into account other characteristics of
the Schwarzchild black hole such as its symmetry and stability.  To do so,
we need to understand how those classical properties can emerge from
a quantum state, described so far in the background independent langauge
of loop quantum gravity. 

Recently, one of us proposed a new perspective on the general problem
of the emergence of particles and other semiclassical states from background
independent approaches to quantum gravity\cite{MP}.  This
makes use of the concept of {\it noiseless subsystems}, developed in the context
of quantum information theory to describe how particle-like states may
emerge there\cite{noisefree}. We shall see in the following that the
black hole problem provides a nice example of the general strategy proposed
in \cite{MP}, while it resolves the present problem\footnote{While this paper was in draft, a paper was posted that reaches a similar conclusion by a different argument \cite{SA}. More recently,
additional arguments for it have been given by \cite{eteradanny}.}.  

\section{Symmetry and noiseless subsystems}

We begin by asking two questions:

\begin{enumerate}

    \item{}How do we find the subspace $\hschw$
    corresponding to a Schwarzchild black hole, as opposed to a
    general surface of a given area that satisfies the appropriate boundary conditions?

    \item{}How do we recognize in that subspace the excitations that,
	 in the classical limit,
    correspond to the quasinormal modes?

\end{enumerate}

We should here make an  important comment, which clarifies the sense in which
these questions are asked. We  note that the boundary conditions we require
for our analysis are more general and apply to all horizons, not just Schwarzchild black holes.  Thus, our framework differs from the
isolated horizon framework\cite{isolated} in one crucial aspect.  
We posit that the states of all quantum black holes
live in a single Hilbert space, so that angular momentum and other multiple
moments are to be measured by quantum operators.  
In the isolated horizon
picture the angular momentum and multiple moments are treated 
classically, and the Hilbert space is defined for each value of them.  Only in the
our framework does it make sense to ask how the states corresponding to
a non-rotating black hole emerge from a more general Hilbert space of
horizon states.

Our two questions are then analogous to problems concerned with the emergence and
stability of  persistent quantum states in condensed matter physics and
quantum information theory.     For the
present purposes, we will use the following idea from the theory of noiseless subsystems in quantum information theory.

We have a complete quantum experiment, which we want to divide into
the system $\cal S$ and environment $\cal E$.  We want to understand
what quantum properties of the system may survive stably in spite of
continual and uncontrollable interactions with the environment.  

The joint Hilbert space decomposes into the product of system and environment,
\f
{\cal H}^{total} = {\cal H}^{\cal S} \otimes  {\cal H}^{\cal E},
\label{eq:decompose}
\ff
while the Hamiltonian  decomposes into the sum
\f
H=H^{\cal S} +  H^{\cal E}  + \hinter
\ff
where $H^{\cal S}$ acts only on the system, $H^{\cal E}$ acts
only on the environment and all the interactions between them
are contained in $H^{int}$.  

The reduced dynamics of ${\cal S}$ is given by a completely positive operator $\phi:{\cal H}^{\cal S}\rightarrow {{\cal H}^{\cal S}}'$, 
\f
\rho^{\cal S}_f=\phi\left[\rho^{\cal S}_i\right]=\tr_{\cal E}\left[U\left(\rho^{\cal S}_i\otimes \rho^{\cal E}_i\right)U^\dagger\right],
\label{eq:phi}
\ff
where the joint state of ${\cal S}$ and ${\cal E}$ evolves unitarily and then the environment is traced out.  While a generic ``noise'' $\phi$ will affect the entire state space ${\cal H}^{\cal S}$, there may be a {\it noiseless subsystem} of ${\cal S}$, and possibly even a subspace of ${\cal H}^{\cal S}$ that is left unchanged by $\phi$, i.e., evolves unitarily.   What follows is a necessary and sufficient condition for the existence of a noiseless subsystem. 

Equation (\ref{eq:phi}) can be rewritten as
\f
\phi\left[\rho^{\cal S}\right]=\sum_k A_k\rho^{\cal S} A_k^\dagger
\ff
where
\f
\langle\psi_i|A_k|\psi_f\rangle=\langle\psi_i |\otimes\langle e_k|U|\psi_f \rangle\otimes |e_k \rangle,
\ff
for any $|\psi_i\rangle, |\psi_f \rangle\in{\cal H}^{\cal S}$ and $\{|e_k\rangle\}$ an  orthonormal basis on ${\cal H}^{\cal E}$.
One can check that $\sum_k A_k^\dagger A_k=1$.  

If ${\cal B}\left({\cal H}^{\cal S}\right)$ is the algebra of all operators acting on ${\cal H}^{\cal S}$, the {\it interaction algebra} $\ainter\subseteq {\cal B}\left({\cal H}^{\cal S}\right)$ is the subalgebra generated by the $A_k$ (assuming that $\ainter$ is closed under $\dagger$ and $\phi$ is unital).  Up to a unitary transformation, $\ainter$ can be written as a direct sum of $d_j\times d_j$ complex matrix algebras, each of which appear with multiplicity $\mu_j$:
\f
\ainter\simeq  \bigoplus_j{\bf 1}_{\mu_j}\otimes{\cal B}\left({\bf C}^{d_j}\right),
\ff
where ${\bf 1}_{\mu_j}$ is the identity operator on ${\bf C}^{\mu_j}$.

The {\it commutant}, $\cainter$ of $\ainter$ is the set of all operators in ${\cal B}\left({\cal H}^{\cal S}\right)$ that commute with every element of $\ainter$, i.e.,
\f
\cainter\simeq\bigoplus_j {\cal B}\left({\bf C}^{\mu_j}\right)\otimes{\bf 1}_{d_j}.
\ff
This decomposition induces a natural decomposition of ${\cal H}^{\cal S}$:
\f
{\cal H}^{\cal S}=\bigoplus_j{\bf C}^{\mu_j}\otimes{\bf C}^{d_j}.
\label{eq:c}
\ff
Note now that any state $\rho$ in $\cainter$ is a fixed point of $\phi$ since it commutes with all the $A_k$:
\f
\phi\left[\rho^{\cal S}\right]=\sum_k A_k\rho^{\cal S} A_k^\dagger=\sum_k A_k A_k^\dagger\rho^{\cal S}=\rho^{\cal S}.
\ff
It can be shown \cite{noisefree} that the reverse also holds, i.e., 
\f
\phi\left[\rho^{\cal S}\right]=\rho\quad\Leftrightarrow\quad\rho\in\cainter.
\ff
Hence, the noiseless subsystem can be identified with the ${\bf C}^{\mu_j}$ in equation (\ref{eq:c}).

These  are relevant for the physical
description of the system, because the interactions with the
environment will not distrurb them and hence are useful for describing the long
term behavior of the system, because they are conserved. 
The relevance of this argument for quantum gravity was proposed in
\cite{MP}. If we divide the quantum state of
the gravitational field arbitrarily into
subsystems, those properties which are conserved under interactions
between the subsystems are going to characterize the low energy limit
of the spacetime geometry.  If classical spacetime physics emerges
in  the low energy
limit, in which these regions are arbitrarily large on
Planck scale,
the commutant of the interaction algebra 
should include the symmetries that characterize classical spacetime in the ground state (presumably the 
Poincar{\'e} or the deSitter group, or some deformation of them).
In the case of vanishing cosmological constant, we then expect that the
low energy limit is characterized by representations of the Poincar{\'e} group.
But these, of course, are the elementary paticles, as described in
quantum field theory on Minkowski spacetime. The noiseless subsystem
idea is then a way to understand how particle states, and the low
energy vacuum could emerge from a background independent quantum
theory of gravity.

In the following section, the system state space is that of the states of loop quantum gravity that correspond to a black hole of a given area.  This is coupled to an environment of bulk spin network states and, while we do not know the quantum dynamics, we do know that a Schwarzchild black hole has a classical SO(3) symmetry.  In the next section, we make the reasonable assumption that there is a microscopic analogue of SO(3) that acts on the system, a symmetry that should be contained in the commutant of the interaction algebra for black holes.

\section{Application to black holes}

In this section we shall look for  the subspace $\hschw$ of states that correspond to a classical Schwarzschild black hole.  We want to show that $\hschw$ is a proper subspace of the space $\hall$ of states of horizons with area $A\pm\Delta A$, i.e.,
\begin{equation}
\hschw \subsetneq \hall.
\end{equation}
The geometry of a horizon is only partially fixed by its area $A$. A Schwarzschild horizon is a very special example of such a geometry that is distinguished by its SO(3) symmetry. Since all the other horizon geometries have quantum counterparts in $\hall$ as well, we expect that the Hilbert space $\hschw$ corresponding to a Schwarzschild black hole is a proper subspace of $\hall$. 

\begin{figure}
\begin{center}
\includegraphics[height=6cm]{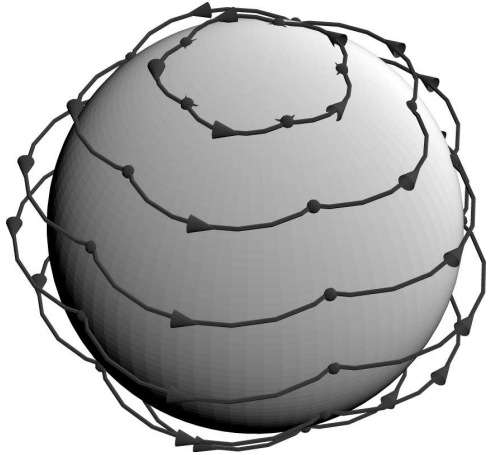}
\end{center}
\caption{The geometry of a quantum horizon is concentrated at a discrete set of punctures. The classical symmetry group SO(3) will thus not act directly on the quantum states representing a Schwarzschild black hole. A discrete group $G_q$, on the other hand, is expected to act on these states. In the classical limit, the action of this discrete group coincides with that of SO(3). For a large number of punctures, the discrete group action shown above would approximate a rotation of the horizon.}\label{fig:action}
\end{figure}

In contrast to the smooth horizon geometry of a classical Schwarzschild black hole, the geometry of the corresponding quantum horizon is concentrated at the punctures and is thus discrete. Consequently, we cannot expect an action of the smooth Lie group SO(3) on $\hschw$. What we can expect is an action of some discrete symmetry group $G_q$ that only on the classical level coincides with the action of SO(3) (see Figure \ref{fig:action}).

What can reasonably be assumed about the group $G_q$?

Since $G_q$ is a group of symmetries we assume it commutes with the action of the Hamiltonian $\hinter$ on the horizon Hilbert space $\hsurf$, i.e.,
\begin{equation}\label{eqn:commute}
G_q \subset \cainter.
\end{equation}
We do not know the precise action of the Hamiltonian $\hinter$ but it is possible to  restrict it sufficiently
to compute the commutant $G_q$.  

We first review the basics of the description of black hole horizons in 
loop quantum gravity\cite{isolated}.  The Hilbert space is at the kinematical level of the form (\ref{eq:decompose}), where the environmental Hilbert space has a basis consisting of all spin networks that end on the horizon, in any number of punctures.  Given that the energy of a black
hole is a function of its area, it makes sense to decompose the  Hilbert space
in eigenspaces of the area operator. The eigenvalues are given by sets of spins at the punctures
$\{ j_i \}$, where $i$ labels the punctures.  Gauss's law acting at the boundary enforces that
the labels on the punctures on the horizon match the spins of the edges of spin network
in the bulk, incident on the horizon.  We then have,
\f
{\cal H}^{total} = \sum_{\{ j_i \} }{\cal H}^{\cal S}_{\{ j_i \} } \otimes  {\cal H}^{\cal E}_{\{ j_i \} },
\label{eq:bhdecompose}
\ff
The environment (sometimes called ``bulk") Hilbert spaces  ${\cal H}^{\cal E}_{\{ j_i \} }$ have  bases consisting
of diffeomorphism classes of spin networks in the bulk, with edges ending on the boundary at the punctures 
with the specified spin labels.  We note that the diffeomorphisms on the surface are fixed.  
(There may also be an exterior boundary at which the diffeomorphisms are also 
restricted.)  

The system, or surface,  Hilbert spaces ${\cal H}^{\cal S}_{\{ j_i \}} $ are direct sums of the one dimensional 
Hilbert spaces of $U(1)$ Chern-Simons
theory on the sphere, with punctures labeled by charges $m_i$ at points $\sigma_i \in S^2$, subject to the conditions
\f
| m_i | \leq j_i  , 
\ff
\f
\sum_i m_i =0  . 
\ff
The level $k$ of each Chern-Simons theory depends
on the $j_i$'s and is given by \cite{isolated},
\f
k= \frac{a_0[\{ j_i \} ]}{4 \pi \gamma}
\ff
where $a_0[\{ j_i \} ] $ is the nearest number to $A[\{ j_i \} ] $ such that the level $k$ is an integer.  We note that
the level $k$ is very large for black holes large in Planck units, so that for some estimates the limit
$k \rightarrow \infty$ can be taken. This is normally done in the computation of the entropy, for which the
difference between the classical and quantum dimensions may be neglected.  

The $U(1)$  connection on the boundary satisfies
\f
F_{ij} (\sigma )= \frac{4 \pi}{k} \sum_i m_i \delta^2 (\sigma, \sigma_i ).
\ff
We then can use a basis of the boundary theory, which is  labeled
\begin{equation}
   \vert (j_1, m_1), (j_2, m_2), \ldots, (j_N, m_N)\rangle .
\label{basis}
\end{equation}

We can now turn to the specification of the interaction algebra. 
We begin by specifying a general interaction algebra, which consists of 
all possible interactions between an environment and a quantum horizon. 
Then we will specialize
to a subalgebra which excludes the case of an
asymmetric environment or external geometry.   The general interaction
algebra, 
$\aginter$, must be generated by operators that
\begin{enumerate}

\item{}Act simultaneously on the environment and system Hilbert space.

\item{}Act locally.   

\end{enumerate}

A sufficient set of generators for $\aginter$ consists of:
\begin{itemize}
\item{} Addition or removal of a puncture.
\item{}  Braidings of the punctures. 
\end{itemize}
The Gauss's law constraint that ties the labels on punctures to the labels on edges of spin networks that
meet them implies that each of these involve changes to both the surface and environment state.  
We note that the braidings generate a group, called the braid group and together with adding and removing punctures, these generate the tangle algebra discussed by Baez in \cite{baez-tangle}.  

Let us consider, for example, a generator in $\aginter$ that corresponds to braiding two punctures, which
we will label $i=1,2$.  It has the effect of rotating, with a positive orientation, the two punctures around each other,
returning them to their original positions. This 
braids the two edges of the bulk spin network, which changes the diffeomorphism class of the bulk state. 

The action of $\hinter$ on $\hsurf$ due to a braiding of the  punctures can be  shown to modify each basis state
(\ref{basis})  by a phase.
As the punctures are unchanged, it is represented by an operator $\hat{\cal B}_{12}$ in ${\cal H}^{\cal S}_{\{ j_i \} }$.
A simple computation in the quantum Chern-Simons theory \cite{CS} shows that this is 
realized by 
\f
\hat{\cal B}_{12} \circ  \vert (j_1, m_1), (j_2, m_2), \ldots, (j_N, m_N)\rangle = e^{\frac{2 \pi \imath (m_1 + m_2}{k} }
 \vert (j_1, m_1), (j_2, m_2), \ldots, (j_N, m_N)\rangle.
\label{eqn:actionofhinter}
\ff

The general interaction algebra does not take into account any symmetries,
hence it describes the general case, in which the black hole horizon 
may have non-vanishing multiple moments\footnote{In the isolated horizon  approach\cite{isolated} angular momentum and multiple moments are coded as classical
parameters.  Here we seek instead
to define different subspaces of the horizon Hilbert space associated with black holes with different macroscopic properties.}.  If we want to specialize
to the case of a black hole with symmetries, we must impose conditions on
the interaction algebra, for a symmetric subsystem cannot exist stably in an asymmetric environment. In principle we could code various kinds of symmetries in the choice of interaction algebra.  But it suffices to consider the simplest
choice, which is the subalgebra $\ainter \subset \aginter$ which consists
of adding and removing punctures plus a symmetric sum of braiding operations,
\f
\hat{\cal B}^T = \sum_{i <j } \hat{\cal B}_{ij}
\ff
where the sum is over all pairs of punctures. 

To restrict the group $G_q$, we now assume that it shares two properties with its classical counterpart SO(3). The first property is:
\begin{description}
\item[P1] The elements of $G_q$ do not change the area of the horizon. 
\end{description}

We take this to mean that the elements of $G_q$ commute with the area 
operator\footnote{In principle we could consider a weaker requirement which 
is that  the elements of $G_q$ do not change the expectation value of the area.
This would be an interesting extension of the usual problem to study.}.
This is trivially true for the action of SO(3) on the classical spacetime and we assume that it is also true for the action of $G_q$. Because of the property \textbf{P1}, the elements of $G_q$ will map the spaces ${\cal H}^{\cal S}_{\{ j_i \} }$ into themselves. A necessary condition for equation (\ref{eqn:commute}) to hold, i.e., for $G_q$ to commute with the action of $\hinter$, is that $G_q$ commutes with the action of $\hinter$ on each of the spaces ${\cal H}^{\cal S}_{\{ j_i \} }$.

On each ${\cal H}^{\cal S}_{\{ j_i \} }$ the braiding generators in  $\hinter$ act as shown in equation 
(\ref{eqn:actionofhinter}). The commutant of $\ainter$ is then easy to find. It is generated by all those operators that just permute the $m_i $ values for given spins $j_i$:
\begin{equation}
\vert \ldots, (j,m_1), \ldots (j,m_K),\ldots \rangle \longrightarrow 
\vert \ldots, (j,m_{\pi(1)}), \ldots (j,m_{\pi(K)}),\ldots \rangle,
\end{equation}
for some $\pi\in  P_K$, the permutation group of $K$ objects. Assuming $G_q$ to be unitary we then have 
\begin{equation}\label{eq:gqpart}
G_q\subset \prod_j P_{K_j},
\end{equation}
where the product is over all the different $j$'s in $\{j_i\}$. 

We now assume that $G_q$ also shares the following property with SO(3):

\begin{description}
\item[P2] $G_q$ acts transitively on the punctures, i.e., for every two punctures there is an element in $G_q$ that connects the two. 
\end{description}

Because of equation (\ref{eq:gqpart}), we know that $G_q$ is a subgroup of 
\begin{equation}
P = \prod_j P_{K_j}.
\end{equation}
For $G_q$ to act transitively, it is necessary that the bigger group $P$ acts transitively on the punctures. Since $P$ is the product of permutation groups, this is only possible if $P$ coincides with just one permutation group. It follows that there is one $j$ such that
\begin{equation}
K_j = N.
\end{equation}
All the punctures have the same spin $j$ and the Hilbert space $\hschw$ consists of those $\hjs$ for which all the $j$'s coincide. It is then straightforward to show that the dimension of the $\hjs$ varies sharply with the spin $j$ and is dominated by the lowest spin $j_{\text{min}}$.

\section{Conclusions}

In this paper, we used a microscopic analogue of the classical SO(3) symmetry of a Schwartzchild 
black hole to restrict the loop quantum gravity state space ${\cal H}_A$ of  black holes of area $A$ to a smaller subspace, left invariant by the symmetry.  The method we used is the noiseless subsystems of quantum information theory, in which the symmetry of the dynamics implies a non-trivial commutant of the interaction algebra of the system.

The construction of the state space ${\cal H}_A$ in loop quantum gravity is a hybrid construction that requires the imposition of black hole horizon conditions at the classical level.  One eventually wants to be able to identify black hole states directly in the quantum theory and derive the classical geometry in the appropriate classical limit.  While the present work has the same hybrid character (we start from ${\cal H}_A$), it indicates that properties of the quantum states can be inferred from the dynamics algebraically, without need for a classical geometry.

This approach also clarifies the important but subtle question of how symmetries can arise from a diffeomorphism invariant state. All the states discussed here
are invariant under spatial diffeomorphisms.  The point is that the commutant,
$G_q$ takes physical, diffeomorphism invariant states, to other distinct physical, diffeomorphism invariant, states. The transitive permutations that we discuss
translate the black hole horizon with respect to the spin networks that define the
external geometry.  This is physically meaningful because these transformations
are not subgroups of the diffeomorphisms, instead they act on the space
of diffeomorphism invariant states. 
Thus, we have  a realization of the proposal in \cite{MP} that in a diffeomorphism invariant theory a symmetry can only arise as a motion of one subsystem with respect to another, and its generators must
live in the commutant of the interaction algebra, defined by the splitting of the
universe into subsystems. 

This proposal has ramifications that might lead to a better
understanding of how the classical properties of black holes arise
from the quantum geometry of the horizon.  
We also see in a simple example, how properties of
classical spacetime geometries can emerge from exact quantum geometries,
by making use of the insights gained from the study of similar questions
in quantum information theory.

We close by mentioning  several queries that may be made regarding the construction 
used here.

\begin{itemize}

\item{} {\it What about the connection to quasi-normal modes? Can we
understand it as other than a coincidence?} It may be possible 
to understand how the quasi-normal modes arise from the states
which transform non-trivially under those generators of the commutant
that become rotations in the $A \rightarrow \infty$ limit. 

\item{}{\it What about rotating black holes?} It is of interest to use these
methods to   characterize the states of a rotating black hole, or a horizon 
with multiple moments, in terms of
the same language.  We expect that this will involve other choices of the
interaction algebra than that made here.  We may note that attempts to
extend the connection between quasi-normal modes and entropy to
rotating black holes have not succeeded; this is something that needs
explanation.  

Note also that while we employ here the same boundary conditions as the isolated 
horizon formalism, the intension is different. In particular, in the isolated horizon picture, rotation and other multiple moments are fixed in the classical description.
The isolated horizon boundary conditions don't change, but the functions on
the phase space that correspond to different observables will depend on
the multiple moments.  In our picture, we seek to recover the generators of
rotation, and hence the conjugate conserved quantities, as operators in a 
single Hilbert space. Hence here, unlike the pure isolated horizon case, we
expect that a single Hilbert space contains the states corresponding to 
black holes with all values of angular momentum.

\item{}{\it What about the states in ${\cal H}_A^{All}$ that are
not in ${\cal H}_A^{Schwarzchild}$? }  These are states whose puctures give
the right area, but are not in the noiseless subsystem we identify.  But they
arose from a quantization of the isolated horizon boundary conditions.

We cannot give a definitive answer without more dynamical input, particularly
a Hamiltonian on the horizon states.   But we can argue that they must correspond to either 
non-static, classical metrics or they correspond to no classical geometry.  

Given that the boundary conditions do not distinguish rotating from non-rotating black holes, some of theses states  must  be associated
with rotating black holes.  Apart from these, it must be that most of these states are 
associated with semiclassical states of non-stationary horizons, corresponding to distorted black holes. These will be excited, transient states of the black hole, which classically decay to stationary states, emitting gravitational
radiation.  How do we know that such states are included in ${\cal H}_A^{All}$?   All that distinguishes 
states in ${\cal H}_A^{All}$ is that there is a horizon of a particular area.  This will include all
such configurations, whether belonging to stationary states, semiclassical states corresponding to non-stationary horizons as well as quantum states that have no semiclassical approximation.  In particular, these must include states corresponding to quasi normal modes.  

Beyond those categories, 
our argument implies that the remaining
states couple strongly to the random noise coming from the environment
of the black hole.  They are then states that cannot play a role in the semiclassical limit.  

The only definitive way to distinguish the different kinds of  states is dynamically.  If we had an effective Hamiltonian for
the horizon and near horizon geometry, then the non-stationary states would have to have higher energy than the stationary states, corresponding to their potential for radiating energy to infinity in gravitational radiation.  Hence, we can deduce from the fact that in the classical theory non-stationary states radiate, that those states would be suppressed in a Boltzman weighted, equilibrium ensemble. In the ensemble ${\cal H}_A^{All}$ they appear unsuppressed, but that is only because no dynamics are invoked, so states of different energies are counted as having the same weight.  

Given that there are many more non-stationary classical configurations than stationary ones, it is reasonable that, for a given horizon area,  those states that
correspond to static black holes must appear to be a minority, as is the case if we compare 
${\cal H}_A^{All}$ with its subspace ${\cal H}_A^{Schwarzchild}$.
But,  it is incorrect to argue just on the basis of the fact that 
${\cal H}_A^{All}$ is larger than ${\cal H}_A^{Schwarzchild}$, that a typical
state corresponding to an equilibrium, static black hole will be in the latter rather than the former, because dynamics has yet to be invoked.    

The point of the construction of this paper is then that, even in the absence of an explicit Hamiltonian, we can use symmetry properties of the Hamiltonian to apply an argument from quantum information to
deduce which subspace of states will emerge in the low energy limit as corresponding to stationary configurations.  The fact that, in the absence of dynamics,
 these are a minority of the ensemble of states with a given area must
be expected and is not an objection against the construction presented here. 

\item{} Finally, we note that the same method could be applied within quantum
cosmology, to study how  states that correspond to symmetric universes 
emerge out of a more general Hilbert space, corresponding to 
the full theory\cite{thankseric}.

\end{itemize}

\section*{ACKNOWLEDGEMENTS}

We are very grateful to Paolo  Zanardi for pointing out an oversight in an earlier version of this paper. We also thank Abhay Ashtekar and Jerzy Lewandowski for very useful discussions clarifying the treatement of symmetries in the isolated horizon formalism.


\begin{thebibliography}{0}

\bibitem{bek1}J.D.\ Bekenstein, ``Black Holes And Entropy'',
{\it Phys.Rev.} D 7, 2333 (1973).

\bibitem{hawking}S.W.\ Hawking,
``Particle Creation By Black Holes'', {\it Commun.Math.Phys.}
43, 199 (1975).

\bibitem{linking}L.\ Smolin, ``Linking topological quantum
field theory and nonperturbative quantum gravity'',
{\it J.Math.Phys.} {\bf 36} (1995) 6417, gr-qc/9505028.

\bibitem{kirill1}K.\ Krasnov, ``On Quantum Statistical
Mechanics of a Schwarzschild Black Hole'' , {\it Gen.Rel.Grav.}\/
{\bf 30} (1998) 53-68, gr-qc/9605047;\\
 C. Rovelli, ``Black hole entropy from loop quantum gravity,"
gr-qc/9603063.

\bibitem{isolated}A.\ Ashtekar, J.\ Baez, K.\ Krasnov, ``Quantum Geometry of
Isolated Horizons and Black
Hole Entropy'',  gr-qc/0005126;\\
  A.\ Ashtekar, J.\ Baez, A.\ Corichi and K.\ Krasnov,
``Quantum geometry and black hole entropy'',
{\it Phys.Rev.Lett.} {\bf 80} (1998) 904-907,  gr-qc/9710007.

\bibitem{olaf-bh}O.\ Dreyer, 
``Quasinormal Modes, the Area Spectrum, and Black Hole Entropy'',
Phys.Rev.Lett. 90 (2003) 081301
 gr-qc/0211076.

\bibitem{imirzi}G. Immirzi, Nucl. Phys. Proc. Suppl. 57, 65 (1997),
gr-qc/9701052.

\bibitem{hod}S. Hod, Phys. Rev. Lett. 81, 4293 (1998), gr-qc/9812002.

\bibitem{motl}L. Motl,
``An analytical computation of asymptotic Schwarzschild quasinormal 
frequencies'',
gr-qc/0212096; \\
L. Motl, A. Neitzke,
``Asymptotic black hole quasinormal frequencies'',
hep-th/0301173.

\bibitem{nomatch}Krzysztof A. Meissner, 
``Black hole entropy in Loop Quantum Gravity'', gr-qc/0407052;\\
Marcin Domagala, Jerzy Lewandowski, 
``Black hole entropy from Quantum Geometry'',
gr-qc/0407051.

\bibitem{SA} S.\ Alexandrov, ``On the counting of black hole states in loop quantum gravity'', gr-qc/0408033.

\bibitem{noisefree} P.\ Zanardi and M.\ Rasetti, {\it Phys.Rev.Lett.}\/{\bf  79}, 3306 (1997);\\
E.\ Knill, R.\ Laflamme and L.\ Viola, {\it Phys.Rev.Lett.}\/{\bf 84}, 2525 (2000);\\
J.\ Kempe, D.\ Bacon, D.A.\ Lidar and K.B.\ Whaley, {\it Phys.Rev.}\/ {\bf A63}, 042307 (2001).

\bibitem{MP}F.\ Markopoulou and D.\ Poulin, in preparation.

\bibitem{CS}D.P.\ Arovas, R.\ Schrieffer, F.\ Wilczek and A.\ Zee, ``Statistical mechanics of anyons'', {\it Nucl.Phys.}\/{\bf B} 251 [FS13] 1985 117;\\
S.\ Rao, ``Anyon primer'', hep-th/9209066.

\bibitem{baez-tangle}J.\ Baez, 
``Quantum Gravity and the Algebra of Tangles'', 
 {\it Class.Quant.Grav.}\/ {\bf 10} (1993) 673-694, hep-th/9205007.  

\bibitem{thankseric}E. Poisson, personal communication.

\bibitem{takashi}T. Tamaki, H.  Nomura, {\it Ambiguity of black hole entropy in loop quantum gravity    }
hep-th/0508142. 

\bibitem{eteradanny}E. R. Livine, D. R. Terno, {\it
Quantum Black Holes: Entropy and Entanglement on the Horizon}, 
gr-qc/0508085.  




\end{thebibliography}
\end{document}